# EFFECT OF SECONDARY IONS ON THE ELECTRON BEAM OPTICS IN THE RECYCLER ELECTRON COOLER *

A. Shemyakin[#], L. Prost, G. Saewert, FNAL, Batavia, IL 60510, U.S.A.

*Abstract*

Antiprotons in Fermilab's Recycler ring are cooled by a 4.3 MeV, 0.1 – 0.5 A DC electron beam (as well as by a stochastic cooling system). The unique combination of the relativistic energy ($\gamma = 9.49$), an Ampere – range DC beam, and a relatively weak focusing makes the cooling efficiency particularly sensitive to ion neutralization. A capability to clear ions was recently implemented by way of interrupting the electron beam for 1-30 μs with a repetition rate of up to 40 Hz. The cooling properties of the electron beam were analyzed with drag rate measurements and showed that accumulated ions significantly affect the beam optics. For a beam current of 0.3 A, the longitudinal cooling rate was increased by factor of ~2 when ions were removed.

## INTRODUCTION

Fermilab's Electron Cooler [1] is used in the 8 GeV Recycler storage ring for storing and preparing antiproton bunches for Tevatron stores. The strength of the longitudinal magnetic field in the cooling section, 105 G, is too low to significantly modify the cooling process. For this case of non-magnetized cooling [2], in practically interesting regimes, the cooling force is approximately proportional to $j_e/\alpha_e^2$, where $j_e$ is the electron current density and $\alpha_e$ is the rms value of the electron angles in the cooling section. $\alpha_e$ can increase if focusing in the beam line is affected by ions created by the electron impact on the residual gas. In this paper, we describe drag rate measurements as a tool to characterize the electron beam quality, estimate the effect of ion accumulation, and present the results of the measurements with ion clearing.

## ELECTRON COOLER AND DRAG RATES

The cooler is based on an electrostatic accelerator, Pelletron [3], working in the energy recovery mode. The DC beam is accelerated in the acceleration tube, is delivered to the cooling section through the so-called supply line, and is returned to the HV terminal through the other Pelletron tube (a mechanical schematic can be found in [4]). The cooling section lacks beam diagnostics that would allow easy measurements of the beam envelope at the interesting range of electron angles ≤ 0.1 mrad. Characterization of the beam properties was accomplished by drag rate measurements [5] of a low-intensity (1-2·$10^{10}$ particles), coasting antiproton beam. Electron cooling decreases the rms radius of the antiprotons in the cooling section to $r_p < 0.5$ mm, noticeably smaller than a typical radius of the electron beam, 2 – 3 mm. As a result, if the two beams propagate parallel to one another with an offset, the drag rate is primarily sensitive to the electron angles in the area that overlaps the antiproton beam. The measurement consists of the following steps: (1) the two beams stay in a concentric position (so called "on-axis") until the antiproton momentum spread reaches an equilibrium; (2) in the cooling section, the electron beam is quickly shifted parallel to the axis; (3) the electron energy is changed by a jump (typically, by 2 keV); (4) for 2 minutes, while the antiproton momentum distribution is dragged toward the new equilibrium point, the average momentum of the antiprotons $\bar{p}$ is measured every ~20 sec; (5) the electron beam position and energy are returned to their initial values; (6) the time derivative $d\bar{p}/dt$ calculated over two minutes is reported as a drag rate. For these conditions, the drag rate is approximately equal to the longitudinal cooling force [6] and is determined primarily by the local value of $j_e/\alpha_e^2$. Because the angles are much more sensitive to variations of focusing than the current density distribution is, the drag rate measurements can be used to estimate changes of the electron angles at a given location.

## SECONDARY IONS

The initial kinetic energy of the secondary ions is close to thermal. The electric field of the electron beam prevents ions from escaping radially, and with no ion clearing mechanisms, the ion density would increase until reaching the electrons' (*i.e.* up to the neutralization factor of $\eta \sim 1$). At $\eta \sim 1$ the focusing effect from ions is a factor of $\gamma^2 \sim 100$ higher than defocusing from the beam space charge. While electron beam envelope simulations with the OptiM code [7] do not agree with measurements quantitatively, qualitatively it predicts that the electron beam space charge becomes important in the beam line at the beam current of $I_e \sim 0.1$ A. Therefore, for the operational range of 0.1– 0.5 A, the effect of ions should be significant at $\eta \sim 1\%$, thus requiring effective ion clearing.

All capacitive pickups monitoring the beam position in the cooler (BPMs) have a negative DC offset on one of its plates, while the other is DC grounded. The resulting electric field prevents the creation of a potential minimum inside the pickup and removes ions in the vicinity of each BPM. To further estimate the process of ion accumulation, we assume the residual gas to be hydrogen at 0.3 nTorr. The calculated time of reaching $\eta \sim 1\%$ is ~200 ms. It is much longer than the time for a thermal – velocity $H_2^+$ ion to fly ~5 m between two neighbouring BPMs, ~3 ms, and, therefore, clearing with the electric field in BPMs should be effective. However, significant size variations of both the electron beam and the vacuum pipe along the beam line create local potential minimums that prevent ions from travelling to the clearing field in the BPMs. Also, solenoidal lenses providing focusing in the beam line are additional barriers for ions. Because the electric field inside the electron beam is primarily radial, the transverse component of the ion velocity is typically much higher than the longitudinal. Due to the conservation of the transverse adiabatic invariant, even the modest magnetic fields of the lenses (≤ 600 G) can reflect the ions. Preliminary analytical estimations [8] had showed that the ion density can reach locally several percents of the electron's, greatly affecting the electron angles. The hope was that the focusing effect of the ion background could be compensated by adjusting the lens settings.

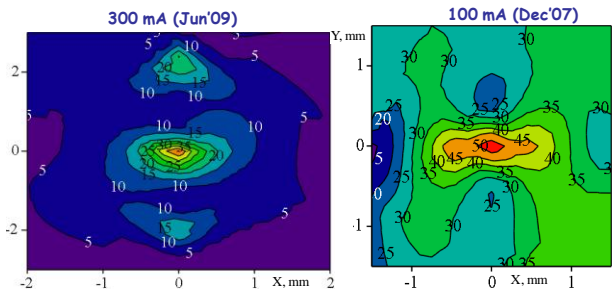

Figure 1: Contour plots of drag rates measured for $I_e = 0.3$ A (left, interpolation from 30 points measured across the beam) and 0.1 A (right, 14 points), both with no ion clearing by interruptions. Contour levels are in MeV/c/hr.

Indeed, the cooling properties of the electron beam were found good enough for what is now a standard operation mode, at $I_e = 0.1$ A. However, cooling efficiency peaked at 0.1 – 0.2 A and decreased at higher currents while it was supposed to be monotonically increasing with $I_e$. Adjusting quadrupoles upstream of the cooling section significantly improved cooling at 0.1 A but did not noticeably change its performance at higher currents [4]. This insensitivity to focusing settings correlates with more recent results of transverse scans of drag rates (Fig.1). While a 0.1 A beam has a relatively smooth cooling profile, at 0.3 A, only three narrow areas provide significant drag rates. This profile corresponds to high-order focusing perturbations that cannot be corrected by adjusting solenoidal lenses and quadrupoles.

## ION CLEARING BY ELECTRON BEAM INTERRUPTIONS

In the potential well created by the electron beam, ions gain the kinetic energy of up to 10 eV (at $I_e = 0.3$ A). Thus, if the electron beam is abruptly turned off, an $H_2^+$ ion reaches the vacuum pipe in 1-2 μs. The capability of interrupting the electron current for 1 – 30 μs was implemented in the electron gun modulator in 2009. Because of schematic specifics, presently, the maximum current from the gun decreases with an increase of the interruption frequency $f_{int}$, $I_{e\_max} = 0.3$ A at $f_{int} = 15$ Hz and 0.1 A at 40 Hz. Tuning in this "ion clearing" mode was done at $I_e = 0.3$ A, $f_{int} = 15$ Hz (while the clearing voltage at BPMs is always on and certainly decreases the ion density, for brevity we refer to the operation with no interruptions as to the mode with no ion clearing). Because of operational limitations, it was done by optimizing cooling of a high-intensity antiproton beam. Specifically, focusing settings were adjusted to minimize the antiproton's equilibrium longitudinal emittance. To avoid overcooling and beam loss due to the resistive wall instability [9], tuning was done with the electron beam shifted in the cooling section by 2.8 mm. The typical antiproton rms radius is 1 mm; therefore, cooling was optimized, first of all, for the periphery of the electron beam.

The most important result for operation is an increase of the longitudinal cooling rate, determined as the time derivative of the rms momentum spread (the measurement procedure is described in [10]). Ion clearing and tuning improved the rate at $I_e = 0.3$ A by a factor of ~2 (Fig.2).

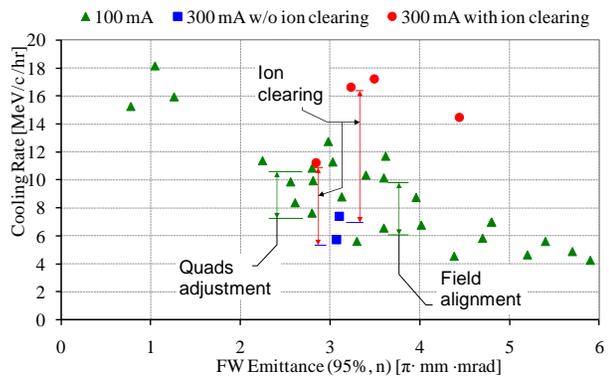

Figure 2. Longitudinal cooling rate measured in 2006-2010. The arrows connect points measured on the same day to show the range of several improvements. All measurements are "on axis".

To understand details of the neutralization effect, several sets of drag rate measurements were performed. First, we determined that varying the interruption width from 2 to 30 μs does not affect the rate. It agrees with the assumption that the contribution to focusing from ions much heavier than hydrogen is insignificant.

The next set of measurements was the dependence of the drag rate on the interruption frequency in two cases: (a) $I_e = 0.3$ A, focusing is optimized at $f_{int} = 15$ Hz; (b) $I_e = 0.1$ A, optimization with no interruptions. The drag rates (Fig.3), measured "on axis", increased toward the frequency for which the corresponding focusing optimization had been made. The results can be compared with the following greatly simplified model.

a. The number of ions $I_e·\eta$ drops to zero at the interruption, increases linearly with time until $t = \tau_c$, and then stays constant, $I_e·\eta_o$ at $t > \tau_c$.
b. Effect of neutralization is equivalent to reducing the beam's space charge by the factor $(1-\eta·\gamma^2)$. Near the axis, it results in an additional electron angle $\Delta\alpha$ proportional to the radial offset $r$, $\Delta\alpha = k·r·I_e·\eta_o·\gamma^2·h(t)$.

Here the function $h(t)$ shows the relative deviation of the neutralization factor at which focusing has been optimized, $h(t) = t/\tau_c$ for $I_e = 0.3$ A and $h(t) = 1-t/\tau_c$ for $I_e = 0.1$ A, and $k$ is a coefficient.

c. In the approximation discussed in the Introduction, the cooling force $F_c$ changes between interruptions as

$$F_c = \frac{F_0}{1+(\Delta\alpha/\alpha_0)^2} \equiv \frac{F_0}{1+(r/a_i)^2 \cdot h(t)^2}, \quad (1)$$

where $\alpha_0$ and $F_0$ are the rms angle and drag force for optimum focusing, and

$$a_i \equiv \frac{\alpha_0}{k \cdot I_e \cdot \eta_0 \cdot \gamma^2}. \quad (2)$$

d. The measured drag rate $F_d$ is the drag force averaged over the period between interruptions and over the antiproton beam cross section. Assuming that antiprotons have a narrow Gaussian distribution $g(r)$ with $r_p \ll$ electron beam size $r_b$, the drag rate is

$$F_d(f_{\text{int}}) = \int_0^\infty g(r) 2\pi r\, dr\, f_{\text{int}} \int_0^{1/f_{\text{int}}} \frac{F_0}{1+(r/a_i)^2 \cdot h(t)^2} dt. \quad (3)$$

Fitting the data with Eq. (3) (solid lines in Fig.3) gives for 0.1 A $\tau_c = 0.4$ s, $a_i = 0.6$ mm and $F_0 = 36$ MeV/c/hr. Because for the 0.3 A set the state with no interruptions could not be reached at constant focusing, the fit is determined by a single parameter $\tau_c \cdot a_i = 0.09$ s·mm (in addition to $F_0 = 36$ MeV/c/hr). For 0.1 A, this product is higher by a factor ~3, in agreement with Eq. (2). At the assumed residual gas composition, $\tau_c = 0.4$ s implies a steady state value $\eta_0$ ~2% and $\langle\eta\rangle$ ~0.2% at 15 Hz.

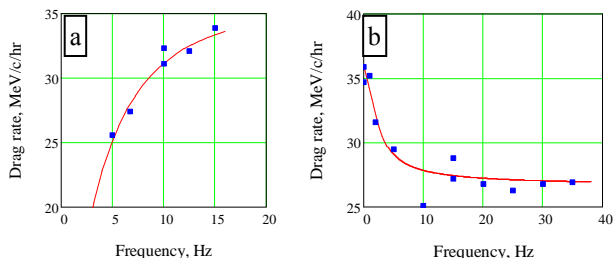

Figure 3. Dependence of the drag rates on the interruption frequency for $I_e = 0.3$ A (left) and 0.1 A (right). The interruption length was 2 µs; beams were on axis. The squares represent the data, and the solid lines are calculations using Eq. (3). Scatter of drag rates at constant conditions is several MeV/c/hr. $r_p \approx 0.5$ mm for 0.3 A measurements and ≈0.3 mm for 0.1 A.

In a separate measurement, the effect of ion clearing was found to be significantly higher further from the axis. While for the 0.1 A set in Fig.3 the ratio of drag rates with and without ion clearing is ~1.4, the measurement at $dY = 1.5$ mm offset gave the ratio of $18/3.5 \approx 5$, not far from the estimation with Eq.(1) $[1 + (dY/a_i)^2] \approx 7$.

Within the model's accuracy (a factor of ~2), the results are self-consistent. In addition, the coefficient $k$ was calculated from OptiM simulations by varying the beam current at constant initial conditions and found to be ~1 rad/A/m. For $\alpha_0 \approx 0.1$ mrad [6], $I_e = 0.1$A, and $\eta_0$ ~2%,

Eq. (2) gives $a_i = 0.5$ mm, close to the experimental result.

With ion clearing, the drag rates were again measured across the electron beam (Fig.4). While the maximum measured rate increased only slightly, the distribution became much smoother and wider. We interpret the asymmetry of the distribution in the case with ion clearing as a result of optimizing the focusing settings while being at an off-axis position.

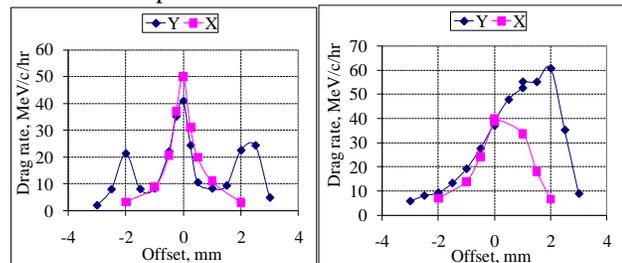

Figure 4. Drag rates measured across the electron beam without (left, part of Fig.1 set) and with ion clearing (right, $f_{int}$ = 15 Hz). $I_e = 0.3$ A.

## SUMMARY

Ions accumulation in the relativistic, DC electron beam of the Recycler cooler significantly affects the cooling properties of the electron beam even at the neutralization factor of ~1%. This effect becomes stronger with an increase of the electron current and, at 0.3 A, results in dramatic nonlinear perturbations that cannot be corrected by adjustment of the solenoidal lenses or quadrupoles.

Interrupting the 0.3 A electron beam for 2 µs at 15 Hz decreases the average neutralization factor by an order of magnitude and allows increasing the cooling rate by a factor of ~2.